\documentclass[%
 reprint,
superscriptaddress,
 amsmath,amssymb,
 aps,
pra,
]{revtex4-2}

\usepackage{graphicx}
\usepackage{dcolumn}
\usepackage{bm}

\begin{document}


\title{Would the fidelity of quantum teleportation be increased by a local filtering operation near a dilaton black hole under decoherence?}

\author{Chun-yao Liu}
\affiliation{%
 College of Physics, Guizhou University, Guiyang 550025, China
}%
\affiliation{%
School of Physics and Electronic Science, Guizhou Normal University, Guiyang 550001, China
}%
 
\author{Zheng-wen Long}%
 \email{zwlong@gzu.edu.cn (corresponding author)}
\affiliation{%
College of Physics, Guizhou University, Guiyang 550025, China
}%


\author{Qi-liang He}
 \email{heliang005@163.com}
\affiliation{%
School of Physics and Electronic Science, Guizhou Normal University, Guiyang 550001, China
}%



\begin{abstract}
Previous studies have shown that the effects of black holes and environmental decoherence generally negatively influence quantum correlations and the fidelity of quantum teleportation in curved spacetime. In our paper, we find that as the dilaton parameter increases, the fidelity of quantum teleportation can either decrease or increase, which suggests that even in the presence of system-environment coupling, the dilaton effect of black hole can positive influence teleportation fidelity; specifically, the dilaton effect can create net fidelity in quantum teleportation under decoherence. This interesting result challenges the long-held belief that the effects of black holes and environmental decoherence can only reduce the fidelity of quantum teleportation. Additionally, we observe an unreported result: if the fidelity of quantum teleportation remains in the classical region, it can be transformed into the quantum region by utilizing a local filtering operation, thereby achieving better fidelity than classical communication. This impressive result may provide new insights for developing an experimental scheme to effectively implement quantum teleportation in the context of dilaton black holes under decoherence. 
\end{abstract}

\maketitle


\section{\label{sec:level1}Introduction}
Quantum teleportation is a fundamental quantum communication task that sends an unknown quantum state from a sender to a receiver using shared quantum entanglement. The theoretical scheme was first proposed by Bennett et al. \cite{a1a} and verified by Bouwmeester et al.'s experiment using single photons \cite{a2a}. In the original protocol, the sender and receiver, who are spatially separated, can only perform local operations and communicate with each other via a classical channel \cite{a3a}.  Quantum teleportation is one of the most important applications of quantum information and has attracted wide attention in recent years  \cite{a5a,a7a,a8a,a9a,a10a,a11a,a12a,a13a,a17a,a18a,a19a,a20a}. 

On the other hand, general relativity and quantum mechanics, the two pillars of modern physics, have combined to give birth to a new theory of relativistic quantum information. In general relativity, black holes are fascinating objects in the universe that are created by the gravitational collapse of a sufficiently massive star. With the development of astronomy, the existence of black holes has been confirmed both indirectly and directly \cite{a21a,a22a,a23a}. Owing to the fascinating properties of black holes, the relativistic quantum investigations near the event horizon of black holes are of fundamental importance. Such studies not only demonstrate practical implications for quantum information protocols \cite{a24a,a25a}, but also provide a better understanding of entropy and the problem of information loss in black holes. Consequently, the behavior of relativistic quantum information in the context of black holes has attracted considerable attention in recent years \cite{a26a,aa26aa,aa27aa,a27a,a28a,a29a,aa28aa,aa29aa,aa30aa,a30a,aaa30aaa}. Many previous studies \cite{aa26aa,aa27aa,aa28aa,aa29aa,aa30aa,aaa30aaa} have shown that the effects of black holes always have a negative impact on quantum correlation and the fidelity of quantum teleportation of Dirac fields in the background of curved spacetime. In addition,  it is well known in black hole physics that the Schwarzschild black hole serves as a paradigmatic example of an uncharged black hole, whereas the Garfinkle-Horowitz–Strominger (GHS) dilaton black hole represents a more general class of static charged black holes \cite{a36a,a37a}. The GHS dilaton black hole arises from gravitational systems coupled to both Maxwell fields and dilaton fields,  and the dilaton effect of the GHS black hole depends on both the black hole's mass and its dilaton field, as the latter constitutes a gravitational source.  The influence of the GHS dilaton black hole's effect on quantum correlations, quantum parameter estimation, entropic uncertainty relation, and quantum steering has been extensively investigated in the theoretical studies \cite{aa34Aaa,aa34Baa,aa34Caa,aa34Daa,aa34Eaa,aa34Faa,aa34Gaa,aa34Haa,aa34Iaa,aa34Jaa,aa34Kaa,aa34Laa}. However, the impact of the dilaton effect in the GHS black hole on the fidelity of quantum teleportation for Dirac fields has not been reported. Therefore, one of our motivations is to investigate whether the dilaton effect of black holes consistently reduces the fidelity of quantum teleportation of Dirac fields.

Most of the studies mentioned above assume an ideal scenario without noise or decoherence. However, it is well known that in the real physical world, quantum systems often inevitably interact with their environment and become open systems. The coupling between the quantum system and the environment causes the initial quantum correlations present in the state of the system to decay \cite{a31a} and also severely reduces the performance of quantum teleportation \cite{a32a}. Therefore, managing decoherence and improving the fidelity of quantum teleportation is essential for performing quantum information tasks, which are the main challenges in achieving quantum information processes \cite{a33a,a34a,a35a}. As a result, another motivation for our research is to discuss the influence of the environment on the fidelity of quantum teleportation within curved spacetime and to explore ways to improve the fidelity of quantum teleportation under environmental influences in the background of black holes.

In this paper, we explore the quantum teleportation of Dirac fields between two users in the background of a GHS dilaton black hole, where Alice’s particle is coupled with the environment. We model the environment using the amplitude damping (AD) channel. The two users, Alice and Bob, initially share an X-type state and employ a standard teleportation scheme (STS) to transfer the unknown state from Alice to Bob. We consider the sender, Alice, remains stationary in an asymptotically flat region, while the receiver, Bob, moves toward the event horizon of the dilaton black hole with uniform acceleration and hovers near the event horizon. Some previous studies \cite{aa26aa,a32a} have found that the fidelity of quantum teleportation decreases as the effects of black holes and environmental influences increase. However, we find that even in the presence of system-environment coupling, the dilaton effect of black hole exerts both negative and positive influences on the fidelity of quantum teleportation. Furthermore, some researchers \cite{a10a,a11a} have also found that in curved spacetime, the fidelity of quantum teleportation of Dirac fields remains in the classical region in some cases, suggesting that the STS scheme cannot provide fidelity better than classical communication. Nevertheless, our results show that we can apply a local filtering operation to shift the fidelity of quantum teleportation from the classical region to the quantum region, thereby achieving better fidelity than classical communication.

The paper is organized as follows. In Section 2, we briefly recall the quantization of Dirac fields in the background of the dilaton black hole beyond single-mode approximation. In Section 3, we discuss the fidelity of quantum teleportation for the X-type state near the event horizon of the dilaton black hole. In Section 4, we examine the influence of the dilaton effect and environmental decoherence on the fidelity of quantum teleportation in dilaton black-hole spacetime. In Section 5, we explore the feasibility of increasing the fidelity of quantum teleportation in dilaton spacetime under decoherence by using a local filtering operation. Finally, a summary and discussion are given in Section 6.

\section{QUANTIZATION OF DIRAC FIELDS IN DILATON BLACK HOLE SPACETIMES}

In this section, we first briefly review the vacuum structure of Dirac particles in GHS dilaton spacetime beyond the single-mode approximation.The metric in the background of a GHS dilation black hole spacetime can be
written as \cite{a36a,a37a} 

\begin{equation}
   {\mathrm{d}}s^2=-(\frac{r-2 M}{r-2 \alpha}) {\mathrm{d}}t^2+(\frac{r-2 M}{r-2 \alpha})^{-1} {\mathrm{d}}r^2+r (r-2 \alpha)   {\mathrm{d}}\Omega^2,
\end{equation}
with $\alpha=Q^2/2M$. Here, $M$, $\alpha$, and $Q$ represent the mass of the black hole, dilaton, and the charge, respectively. For simplicity, we set $G=c=\hbar=\kappa_{B}=1$  throughout this paper. Furthermore, the dilaton $\alpha$ and the mass $M$ of the black hole should satisfy $\alpha<M$.

In a general background of curved spacetime, the massless Dirac equation can be written as \cite{a38a,a39a}
\begin{equation}
    [\gamma^{a}  e_{a}^{\mu}(\partial_{\mu}+\Gamma _{\mu})]  \Psi = 0,
\end{equation}
where $\gamma^{a}$ are the Dirac–Pauli matrices, the four-vectors $e_{a}^{\mu}$ represent the inverse of the tetrad defined by $g_{\mu \nu }=\eta _{a b} e^{a}_{\mu} e^{b}_{\nu}$ with $\eta _{a b}=diag(-1,1,1,1)$, and $\Gamma _{\mu}$ is the spin connection coefficient, given by $\frac{1}{8} [\gamma^{a},\gamma^{b}] e_{a}^{\nu} e_{b \nu ; \mu}$. 

To separate the Dirac equation, we utilize a tetrad as
\begin{equation}
e^{a}_{\mu}=diag(\sqrt{f'},\frac{1}{\sqrt{f'}},\sqrt{r \tilde{r}},\sqrt{r \tilde{r}} \sin \theta),
\end{equation}
with $f'=(r-2M)/\tilde{r}$, and $\tilde{r}=r-2 \alpha $. Then in the GHS dilaton spacetime, Eq.(2) turns into
\begin{eqnarray}
&-\frac{\gamma_{0}}{\sqrt{f'}} \frac{\partial\Psi  }{\partial t} +\gamma_{1} \sqrt{f'} [\frac{\partial }{\partial r}+\frac{r-\alpha}{r \tilde{r}}+\frac{1}{4 f'} \frac{ \mathrm{d} f'}{ \mathrm{d} r}] \Psi \nonumber\\
&+\frac{\gamma_2}{\sqrt{r \tilde{r}}} [\frac{\partial }{\partial \theta }
+\frac{\cot \theta}{2}] \Psi+\frac{\gamma_3}{\sqrt{r \tilde{r}} \sin \theta } \frac{\partial \Psi }{\partial \varphi}=0. 
\end{eqnarray}

If we rewrite $\Psi $  as  $\Psi =  f'^{-\frac{1}{4}} \Phi$ and use an ansatz for the Dirac spinor similar to Ref. \cite{a40a}, the solution of the Dirac equation near the event horizon can be obtained. For the regions outside and inside the event horizon, the positive frequency of outgoing solutions can be expressed as \cite{a41a,a42a}
\begin{eqnarray}
\Psi _{out,k}^{+}=\mathcal{G} e^{- i \omega \mathcal{U}}, \Psi _{in,k}^{+}=\mathcal{G} e^{ i \omega \mathcal{U}},     
\end{eqnarray}
where $\mathcal{U}=t-r_{*}$ is the tortoise coordinate, $\mathcal{G}$ being a four-component Dirac spinor, and $k$ is the wave vector labeling the modes hereafter. With the help of Eq.(5), the Dirac field can be expanded as
\begin{eqnarray}
\Psi&=\int\mathrm{d}k[\hat{a}^{out}_{k} \Psi^{+}_{out,k}+
\hat{b}^{out \dagger}_{-k} \Psi^{-}_{out,k}\nonumber\\
&+\hat{a}^{in}_{k} \Psi^{+}_{in,k}+\hat{b}^{in \dagger}_{-k} \Psi^{-}_{in,k}],
\end{eqnarray}
where $\hat{a}^{in}_{k}$ and $\hat{b}^{in \dagger}_{k}$  are the fermion annihilation and antifermion creation operators acting on the quantum state of the interior region of the black hole, the same for $\hat{a}^{out}_{k}$ and $\hat{b}^{out \dagger}_{k}$ acting on the state of the exterior region of the black hole. The annihilation and creation operators satisfy the usual anticommutation relations $\left \{\hat{a}^{out}_{k},\hat{a}^{out }_{{k}'}\right \}=\delta _{k {k}'}$ and $\left \{\hat{a}^{out}_{k},\hat{a}^{out \dagger}_{{k}'}\right \}=\left \{\hat{a}^{out \dagger}_{k},\hat{a}^{out \dagger}_{{k}'}\right \}=0$. 

Following the suggestion of Damour and Ruffini \cite{a41a}, we perform an analytic continuation of Eq.(5) to obtain a complete basis of positive energy modes (Kruskal modes). However, in Ref. \cite{a43a}, Bruschi et al. revealed that the single-mode approximation, commonly employed in relativistic quantum information scenarios, exhibits limitations and is not applicable to the analysis of quantum properties for arbitrary quantum states. Since the Kruskal observer possesses the freedom to generate excitations within any accessible mode, the single-frequency Kruskal mode cannot be directly mapped to a set of single-frequency modes. To avoid this inconsistency, we adopt the Unruh mode \cite{a43a,aa43Aaa,aa43Baa,aa43Caa,aa43Daa}, which extends beyond the single-mode approximation for quantization of the Dirac field and has been widely applied in relativistic quantum information research. Notably, the study \cite{aa43Caa} further demonstrates that this theoretical approach can help us to analyze a more general family of maximally entangled fermionic states, for which the single-mode approximation does not hold. The method based on the Unruh mode is grounded in the techniques established in prior research \cite{a39a,a40a,a41a,a42a}. Specifically, we construct a set of distinct creation operators, which are formulated as linear combinations of creation operators derived from both the inner and outer regions, which take the forms as
\begin{eqnarray}
&\tilde{c}_{k,R}&=\cos r \hat{a}^{out}_{k}-\sin r \hat{b}^{in \dagger}_{-k},\nonumber\\
&\tilde{c}_{k,L}&=\cos r \hat{a}^{in}_{k}-\sin r \hat{b}^{out \dagger}_{-k},\nonumber\\
&\tilde{c}_{k,R}^{\dagger}&=\cos r \hat{a}^{out \dagger}_{k}-\sin r \hat{b}^{in}_{-k},\nonumber\\
&\tilde{c}_{k,L}^{\dagger}&=\cos r \hat{a}^{in \dagger}_{k}-\sin r \hat{b}^{out}_{-k},
\end{eqnarray}
where $\cos r=[e^{-8 \pi \omega (M-\alpha)}+1]^{-\frac{1}{2}}$ and $\sin r=[e^{8 \pi \omega (M-\alpha)}+1]^{-\frac{1}{2}}$, the subscripts $L$ and $R$ are labeled as left and right modes, respectively. After properly normalizing the state vector, the Kruskal vacuum is found to be $\left | 0 \right \rangle _{k}=\otimes_{k} \left | 0_{k} \right \rangle _{K}=\otimes_{k} \left | 0_{k} \right \rangle _{R} \otimes_{k} \left | 0_{k} \right \rangle _{L}.$ Using the operators ordering $\hat{a}^{out}_{k}\hat{b}^{in}_{-k}\hat{b}^{out}_{-k}\hat{a}^{in}_{k}$ in Eq.(7), the Kruskal vacuum state $\left | 0_{k} \right \rangle _{K}$ for mode $k$ can be expressed as \cite{a43a}
\begin{eqnarray}
\left | 0_{k} \right \rangle _{K}&=&(\tilde{c}_{k,R} \otimes \tilde{c}_{k,L}) \left | 0_{k} \right \rangle _{R} \otimes \left | 0_{k} \right \rangle _{L}\nonumber\\ &=&(\cos r \left | 0_{k} \right \rangle_{out}^{+} \left | 0_{-k} \right \rangle _{in}^{-}+\sin r \left | 1_{k} \right \rangle_{out}^{+} \left | 1_{-k} \right \rangle _{in}^{-}) \nonumber\\ &&\otimes (\cos r \left | 0_{-k} \right \rangle_{out}^{-} \left | 0_{k} \right \rangle _{in}^{+}-\sin r \left | 1_{-k} \right \rangle_{out}^{-} \left | 1_{k} \right \rangle _{in}^{+})\nonumber\\ &=&\cos^{2} r\left | 0000 \right \rangle-\sin r \cos r \left | 0011 \right \rangle\nonumber\\ &&+\sin r \cos r \left | 1100 \right \rangle-\sin^{2} r\left | 1111 \right \rangle,
\end{eqnarray}
where $\left |m n {m}' {n}' \right \rangle=\left | m_{k} \right \rangle_{out}^{+} \left | n_{-k} \right \rangle_{in}^{-} \left | {m}'_{-k} \right \rangle_{out}^{-} \left | {n}'_{k} \right \rangle_{in}^{+}$. The sets $\left \{ \left | n_{k} \right \rangle   _{out}^{+}   \right \} $ and $\left \{ \left | n_{-k} \right \rangle   _{in}^{-}   \right \} $  are the orthonormal bases for the exterior and interior regions of the dilaton black hole, respectively, and the $\left \{ +,- \right \} $ superscript on the kets indicates the fermion and antifermion vacua. For the observer Bob, who hovers near the event horizon, the spectrum of Hawking radiation detected by his instrument is given by $N^{2}_{\omega }= _{K}\left\langle  0  \right|  \hat{a}_{k}^{out\dagger} \hat{a}_{k}^{out}\left | 0 \right \rangle _{K}=\frac{1}{e^{\frac{\omega }{T}}+1}$, with $T=\frac{1}{8 \pi (M-\alpha )} $ represents the Hawking temperature of the black hole, which indicates that observer Bob detects a thermal Fermi–Dirac distribution of particles. Due to the Pauli exclusion principle, only the first excited state for each fermion mode $\left | 1_{k} \right \rangle_{K}^{+}$ is permitted, and the same applies to antifermions. The first excited state for the fermion mode can be written as 
\begin{eqnarray}
\left | 1_{k} \right \rangle _{K}^{+}&=&[q_{R} (\tilde{c}_{k,R}^{\dagger} \otimes I_{L})+q_{L} (I_{R} \otimes \tilde{c}_{k,L}^{\dagger})] \left | 0_{k} \right \rangle _{R} \otimes \left | 0_{k} \right \rangle _{L}\nonumber\\ &=& q_{R}[\cos r \left | 1000 \right \rangle-\sin r \left | 1011 \right \rangle]\nonumber\\ && +q_{L}[\sin r \left | 1101 \right \rangle+\cos r \left | 0001 \right \rangle],
\end{eqnarray}
where $\left | q_{R}  \right |^{2} +\left | q_{L}  \right |^{2}=1,$ $\left | q_{R}  \right |^{2}$ and $\left | q_{L}  \right |^{2}$ denote the probability of right mode and left mode, respectively, which means that particles and antiparticles can radiate randomly toward both the interior and exterior regions from the event horizon.

\section{THE FIDELITY OF QUANTUM TELEPORTATION FOR X-TYPE STATE}
In this section, we present a short overview of X-type states, the general principles of quantum teleportation in the Dirac field, utilizing the fully entangled fraction of the bipartite state as a basis. Here, we consider the universally common X-type state of the bipartite system, which can be expressed in terms of its density matrix in the basis $\left \{\left | 00 \right \rangle,\left | 01 \right \rangle, \left | 10 \right \rangle,\left | 11 \right \rangle\right \} $ as follows \cite{a10a,a11a}
\begin{equation}
\rho_{X}=\left( \begin{array}{cccc}
 \rho_{11} & 0 & 0 &-\rho_{14} \\
  0& \rho_{22} & -\rho_{23} & 0\\
  0& -\rho_{23} & \rho_{33} & 0\\
 -\rho_{14} & 0 &0  &\rho_{44}
\end{array} \right),
\end{equation}
Eq.(10) characterizes a valid quantum state, given that the conditions of unit trace and positivity are met: $\rho_{11}+\rho_{22}+\rho_{33}+\rho_{44}=1$, $\rho_{22}+\rho_{33}\ge \left | \rho_{23} \right |^{2}$, and $\rho_{11}+\rho_{44}\ge \left | \rho_{14} \right |^{2}$. If the condition $\rho_{22} \rho_{33}< \left | \rho_{14} \right |^{2}$ or $\rho_{11} \rho_{44}< \left | \rho_{23} \right |^{2}$ is satisfied, then the X-type state is considered to be entangled.

In a general teleportation scheme related to curved spacetime, Alice and Bob first share an X-type state $(\rho_{X})$ in an asymptotically flat region. The unknown pure state that Alice intends to teleport to Bob is denoted by $\left | \phi  \right \rangle$. Then, some trace-preserving and local quantum operations and classical communication (LOCC) operations are employed by Alice and Bob to be applied to their respective systems. Following these operations, the final state of Bob is expressed as
$\rho_{B}=Tr_{A,C} [M (\left | \phi   \right \rangle \left\langle\phi \right| \otimes \rho_{X})],$ where $M$ represents the trace-preserving LOCC operation. Therefore, the fidelity of quantum teleportation, which serves as a measure of the quality of the teleportation process, can be read as \cite{a44a}
\begin{equation}
F=\left\langle\phi \right|\rho_{B} \left | \phi   \right \rangle =\frac{f d+1}{d+1}, 
\end{equation}
where $d$ denotes the the dimension of Hilbert space $H_{A} \otimes H_{B}=C^{d} \otimes C^{d} $ 
and $f$ is the fully entangled fraction. The achievable fidelity of teleportation in the STS is uniquely determined by the fully entangled fraction of the bipartite state, which is defined as \cite{a45a}
\begin{equation}
f(\rho)=\max_{\varphi} \left\langle\varphi \right|\rho \left |\varphi \right \rangle, 
\end{equation}
with $\left | \varphi  \right \rangle$ is taken over all the maximally entangled states. The STS necessitates that the X-type state satisfies the condition $f>1/d$ to achieve better fidelity than classical communication, which corresponds to the quantum region. In this paper, we will focus exclusively on this state within this region. 

If the density matrix elements of the X-type state in Eq.(10) meet the conditions $\rho_{22} +\rho_{33} \ge 1/2$ and $\rho_{23} \ge \frac{1}{2} (1-\rho_{22}-\rho_{33})$ , the fully entangled fraction can be written as
\begin{equation}
f(\rho_{X})=\frac{1}{2} (\rho_{22} +\rho_{33}+2 \rho_{23}) \ge \frac{1}{2}.
\end{equation}
Since we are focusing solely on the quantum region where $f>1/d=1/2$, the above conditions will be assumed for the X-type state in the following discussion.

\section{DILATON EFFECT ON FIDELITY OF QUANTUM TELEPORTATION IN DILATON SPACETIME UNDER DECOHERENCE}

Initially, we assume that Alice and Bob share an X-type state \cite{a10a,a11a}, as described in Eq. (10), at the same point in the asymptotically flat region of the dilaton black hole. Alice remains stationary in this region and interacts with a dissipative environment, which can be modeled by the well-known amplitude damping (AD) channel \cite{a47a}. Meanwhile, Bob hovers close to the event horizon of the dilaton black hole, where he detects a thermal Fermi-Dirac distribution of particles. The interaction of Alice’s state with the environment can be described by \cite{a47a}
\begin{eqnarray}
&&\left | 0  \right \rangle _{A}\left | 0  \right \rangle _{E}\longrightarrow \left | 0  \right \rangle _{A}\left | 0  \right \rangle _{E},\nonumber\\
&&\left | 1  \right \rangle _{A}\left | 0  \right \rangle _{E}\longrightarrow \sqrt{1-p} \left | 1  \right \rangle _{A}\left | 0  \right \rangle _{E}+\sqrt{p} \left | 0  \right \rangle _{A}\left | 1  \right \rangle _{E},
\end{eqnarray}
where $0\le p \le 1$ denotes the magnitude of the decoherence. According to Eqs. (8), (9), and (14), we can rewrite Eq.(10). Since the region outside the event horizon is causally isolated from the region inside it, and assuming that Bob's detector is only sensitive to particle modes, we trace over the state of the interior region, the environment, and the antiparticle mode in the exterior region, resulting in the following X state (see Appendix A for an explicit derivation)
\begin{equation}
\rho_{X}^{A B^{+}_{out}}=\left( \begin{array}{cccc}
 \rho_{11}^{A B^{+}_{out}} & 0 & 0 &-\rho_{14}^{A B^{+}_{out}} \\
  0& \rho_{22}^{A B^{+}_{out}} & -\rho_{23} ^{A B^{+}_{out}}& 0\\
  0& -\rho_{23}^{A B^{+}_{out}} & \rho_{33}^{A B^{+}_{out}} & 0\\
 -\rho_{14}^{A B^{+}_{out}} & 0 &0  &\rho_{44}^{A B^{+}_{out}}
\end{array} \right),
\end{equation}
where
\begin{eqnarray}
\rho_{11}^{A B^{+}_{out}}&=&\cos^{2}r (\rho_{11}+q^{2}_{L} \rho_{22})+p\cos^{2}r (\rho_{33}+ q^{2}_{L} \rho_{44}),  \nonumber\\
\rho_{22}^{A B^{+}_{out}}&=&\sin^{2}r \rho_{11}+\cos^{2}r q^{2}_{R} \rho_{22}+\sin^{2}r \rho_{22} \nonumber\\
&&+p(\sin^{2}r \rho_{33}+\cos^{2}r q^{2}_{R} \rho_{44}+\sin^{2}r \rho_{44}),  \nonumber\\
\rho_{33}^{A B^{+}_{out}}&=&(1-p) (\cos^{2}r \rho_{33}+\cos^{2}r q^{2}_{L} \rho_{44}), \nonumber\\
\rho_{44}^{A B^{+}_{out}}&=&(1-p) (\sin^{2}r \rho_{33}+\cos^{2}r q^{2}_{R} \rho_{44}+\sin^{2}r \rho_{44}), \nonumber\\
\rho_{14}^{A B^{+}_{out}}&=&\sqrt{1-p} \cos r q_{R} \rho_{14}, \nonumber\\
\rho_{23}^{A B^{+}_{out}}&=&\sqrt{1-p} \cos r q_{R} \rho_{23},
\end{eqnarray}
with the $\cos r$ and $\sin r$ defined in Sec.2. We assume that the state $\rho_{X}^{A B^{+}_{out}}$ satisfies the condition 
\begin{eqnarray*}
    \rho_{22}^{A B^{+}_{out}}+\rho_{33}^{A B^{+}_{out}}\ge \frac{1}{2}.
\end{eqnarray*}

Thus, we can obtain fully entangled fraction as
\begin{eqnarray}
&f(\rho_{X}^{A B^{+}_{out}})&=\frac{1}{2} (\sin^{2}r \rho_{11}+\cos^{2}r q^{2}_{R} \rho_{22}+\sin^{2}r \rho_{22}  \nonumber\\
&&+p(\sin^{2}r \rho_{33}+\cos^{2}r q^{2}_{R} \rho_{44}+\sin^{2}r \rho_{44}) \nonumber\\
&&+ (1-p) (\cos^{2}r \rho_{33}+\cos^{2}r q^{2}_{L} \rho_{44})\nonumber\\
&&+2 \sqrt{1-p} \cos r q_{R} \rho_{23} ).
\end{eqnarray}

The change of $f(\rho_{X}^{A B^{+}_{out}})$ related to the dilaton can be give by

\begin{eqnarray*}
&\Delta_{\alpha}& f(\rho_{X}^{A B^{+}_{out}}(\alpha_{0}))=f(\rho_{X}^{A B^{+}_{out}})|_{(\alpha=\alpha_{0})}-f(\rho_{X}^{A B^{+}_{out}})|_{(\alpha=0)}  \nonumber\\ &&=\frac{1}{2} \left \{2 \sqrt{1-p} q_{R} \rho_{23} [(e^{-8 \pi \omega (M-\alpha_{0})}+1)^{-\frac{1}{2}} \right. \nonumber\\&&-(e^{-8 \pi \omega M}+1)^{-\frac{1}{2}}]       
+[-\rho_{11}- \left | q_{L}  \right |^{2} \rho_{22} \nonumber\\
&&+  p \left | q_{R}  \right |^{2} \rho_{44} -p(\rho_{33}+\rho_{44}) \nonumber\\
&&+ (1-p)  (\rho_{33}+ \left | q_{L}  \right |^{2} \rho_{44} )]  (e^{-8 \pi \omega (M-\alpha_{0})}+1)^{-1} \nonumber\\
&&+[ e^{8 \pi \omega M} (\rho_{11}+ \left | q_{L}  \right |^{2} \rho_{22} -(1-p) (\rho_{33}+ \left | q_{L}  \right |^{2} \rho_{44}) \nonumber\\ 
&& \left. +p(\rho_{33}+\rho_{44}-\left | q_{R}  \right |^{2} \rho_{44}))] (e^{8 \pi \omega M}+1)^{-1}  \right\}.
\end{eqnarray*}

\begin{figure}
\begin{center}
\includegraphics[width=6cm]{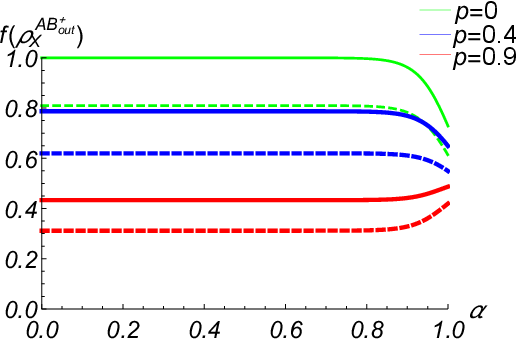}
\caption{\label{fig:fig1} 
The fully entangled fraction $f(\rho_{X}^{A B^{+}_{out}})$ as a function of the dilaton parameter $\alpha$ for different decoherence parameters $p$ and $q_{R}$. The initial parameters are set as: $\rho_{11}=\rho_{44}=\rho_{14}=0$ and  $\rho_{22}=\rho_{33}=\rho_{23} = \frac{1}{2}$. The green, blue, and red curves correspond to $p=0,0.4,0.9$, respectively. The solid line denotes $q_{R}=1$ and the dashed line denotes $q_{R}=0.8$.}
\end{center}
\end{figure}

From the above equation, we can clearly observe the dependence of $f(\rho_{X}^{A B^{+}_{out}})$ on the dilaton parameter $\alpha$, that is, $\Delta_{\alpha} f(\rho_{X}^{A B^{+}_{out}}(\alpha_{0}))>0$ denotes $\frac{\partial f(\rho_{X}^{A B^{+}_{out}})}{\partial \alpha} \mid_{\alpha_{0}}>0$, which means that $f(\rho_{X}^{A B^{+}_{out}})$ increases with the dilaton parameter $\alpha$, $\Delta_{\alpha} f(\rho_{X}^{A B^{+}_{out}}(\alpha_{0}))<0$ denotes $\frac{\partial f(\rho_{X}^{A B^{+}_{out}})}{\partial \alpha} \mid_{\alpha_{0}}<0$, which indicates that $f(\rho_{X}^{A B^{+}_{out}})$ decreases with the dilaton parameter $\alpha$.

In Fig.1, we demonstrate the dynamical behavior of the fully entangled fraction $f(\rho_{X}^{A B^{+}_{out}})$ as a function of dilaton parameter $\alpha$ when Alice is coupled to a dissipative environment. From the green line and the blue line, it can be observed that when the decoherence parameter p is relatively small, the fully entangled fraction $f(\rho_{X}^{A B^{+}_{out}})$ decreases with the increase in the dilaton parameter $\alpha$, which indicates that the dilaton effect of black holes exerts a detrimental influence. In contrast, from the red line of Fig.1, we can find that when the decoherence parameter $p$ is relatively large, the fully entangled fraction $f(\rho_{X}^{A B^{+}_{out}})$ increases as the dilaton parameter $\alpha$ increases. This means that the influence of the dilaton effect on the fully entangled fraction is not always negative when Alice’s qubit is coupled with the environment. Additionally, we can also see that the fully entangled fraction always decreases as the parameter $q_{R}$ goes down. This shows that we can choose the optimal Unruh mode with $q_{R}=1$ to achieve the best transmission of an unknown pure state between Alice and Bob.

\begin{figure}
\begin{center}
\includegraphics[width=6cm]{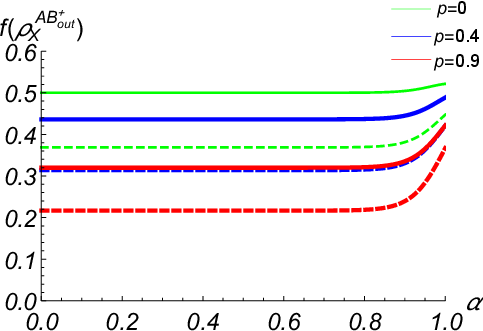}
\caption{\label{fig:fig2} 
 The fully entangled fraction $f(\rho_{X}^{A B^{+}_{out}})$ as a functions of the dilaton parameter $\alpha$ for different decoherence parameters $p$ and $q_{R}$. The initial parameters are set as: $\rho_{11}=\sqrt{2}-1$, $\rho_{22}=\frac{1}{2}$, $\rho_{33}=\frac{3-2 \sqrt{2}}{2}$, $\rho_{14}=\rho_{44}=0$ and $\rho_{23}=\frac{\sqrt{2}-1}{2}$. The green, blue, and red curves correspond to $p=0,0.4,0.9$, respectively. The solid line denotes $q_{R}=1$ and the dashed line denotes $q_{R}=0.8$.}
\end{center}
\end{figure}

\begin{figure}
\begin{center}
\includegraphics[width=6cm]{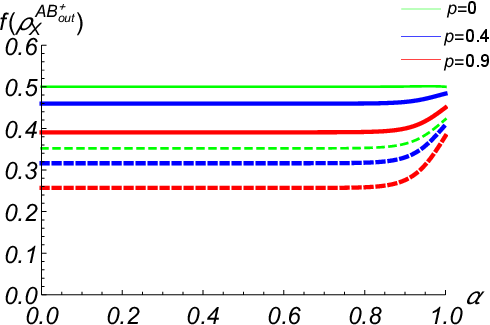}
\caption{\label{fig:fig3} 
The fully entangled fraction $f(\rho_{X}^{A B^{+}_{out}})$  as a functions of the dilaton parameter $\alpha$ for different decoherence parameters $p$ and $q_{R}$. The initial parameters are set as: $\rho_{11}=\frac{\sqrt{2}-1}{2}$, $\rho_{22}=\frac{\sqrt{2}}{2}$, $\rho_{33}=\frac{3-2 \sqrt{2}}{2}$, $\rho_{14}=\rho_{44}=0$ and $\rho_{23}=\frac{\sqrt{2}-1}{4}$. The green, blue, and red curves correspond to $p=0,0.4,0.9$, respectively. The solid line denotes $q_{R}=1$ and the dashed line denotes $q_{R}=0.8$.}
\end{center}
\end{figure}

In Fig. 2, the variation of the fully entangled fraction is plotted as a function of the dilaton parameter $\alpha$, when the initial state parameters are chosen as $\rho_{11}=\sqrt{2}-1$, $\rho_{22}=\frac{1}{2}$, $\rho_{33}=\frac{3-2 \sqrt{2}}{2}$, $\rho_{14}=\rho_{44}=0$ and $\rho_{23}=\frac{\sqrt{2}-1}{2}$. We can find that as the dilaton parameter $\alpha$ increases, the fully entangled fraction $f(\rho_{X}^{A B^{+}_{out}})$ increases. This indicates that the dilaton effect of the black hole positively influences the fully entangled fraction. Consequently, the dilaton effect can enhance the net fidelity of quantum teleportation. Furthermore, we also observe that the fully entangled fraction $f(\rho_{X}^{A B^{+}_{out}})$ increases with elevation of $q_{R}$, which means that the fully entangled fraction is influenced by the choice of Unruh modes, with the Unruh mode at $q_{R}=1$ consistently being optimal for achieving the highest fully entangled fraction between Alice and Bob. Therefore, these results have the potential to provide valuable insights for a more comprehensive understanding of the effects of the dilaton black hole and decoherence on the efficient execution of quantum information tasks in real quantum systems. It is worth emphasizing that how to suppress the decoherence effect and improve the fidelity of quantum teleportation in a real system operating within a curved space-time framework is worth studying.

In Fig. 3, we plot the fully entangled fraction as a function of the dilaton parameter
$\alpha$ when we set the parameters of the initial state as $\rho_{11}=\frac{\sqrt{2}-1}{2}$, $\rho_{22}=\frac{\sqrt{2}}{2}$, $\rho_{33}=\frac{3-2 \sqrt{2}}{2}$, $\rho_{14}=\rho_{44}=0$ and $\rho_{23}=\frac{\sqrt{2}-1}{4}$. From the green solid line in Fig. 3, we can see that when $q_{R}=1$ and $p=0$, the fully entangled fraction is almost unaffected by the dilaton effect and remains constant around 0.5. Furthermore, we also observe that when the system becomes coupled with the environment, the fully entangled fraction increases with the growth of the dilaton parameter $\alpha$. This indicates that the dilaton effect of the black hole has a positive influence on the fully entangled fraction. In particular, as shown by the red dashed line in the figure, the magnitude of the increase in the fully entangled fraction is greatest with increasing dilaton parameter $\alpha$. This implies that at $q_{R}=0.8$ and $p = 0.9$, the dilaton effect of the black hole can create more net fidelity of quantum teleportation. Additionally, the figure clearly shows that as the decoherence parameter $p$ increases and $q_{R}$ decreases, the fully entangled fraction also decreases. This indicates that in non-optimal Unruh modes and under decoherence, finding ways to enhance the fidelity of quantum teleportation becomes very interesting.

\section{ENHANCEMENT OF FIDELITY OF QUANTUM TELEPORTATION IN DILATON SPACETIME UNDER DECOHERENCE BY LOCAL FILTERING OPERATION}

In this section, we investigate the feasibility of increasing the fidelity of quantum teleportation in dilaton spacetime under decoherence using a local filtering operation. The local filtering operation is defined by a non-trace preserving operator and can recover and increase entanglement to some degree \cite{a48a,a49a,a50a}. According to Ref. \cite{a48a}, the local filtering operation can be expressed as
\begin{equation}
M_{f t}=\left( \begin{array}{cccc} 
\sqrt{1-ft} & 0  \\  
0& \sqrt{ft} 
\end{array} \right),
\end{equation}
with $ft$ denoting the strength of the filtering operation. Initially, we consider that Alice and Bob share an X-type state at the same initial point in flat Minkowski spacetime, which can be described in Eq.(10). Subsequently, Alice passes her particle through the AD decohering channel and performs a local filtering operation on it. Meanwhile, Bob freely falls toward a dilaton black hole and hovers outside the event horizon. Here, we only performed the local filtering operation on Alice, as it is easier to implement in practice than to apply the operation to both Alice and Bob. By making use of the Eqs. (15) and (18), the density matrix of the physically accessible system $\rho_{f t}^{' A B^{+}_{out}}$ can be written as 
\begin{eqnarray}
&&\rho_{f t}^{'A B^{+}_{out}}=\frac{ \left ( M_{f t}\otimes {\uppercase\expandafter{\romannumeral1}_{2}}  \right ) \rho_{X}^{ A B^{+}_{out}} \left ( M_{f t}\otimes {\uppercase\expandafter{\romannumeral1}_{2}}  \right ) ^{\dagger}} {Tr\left [  \left ( M_{f t}\otimes {\uppercase\expandafter{\romannumeral1}_{2}}  \right ) \rho_{X}^{ A B^{+}_{out}} \left ( M_{f t}\otimes {\uppercase\expandafter{\romannumeral1}_{2}}  \right ) ^{\dagger}              \right ] }  \nonumber\\
&& =\left( \begin{array}{cccc}
 \rho_{11}^{'A B^{+}_{out}} & 0 & 0 &-\rho_{14}^{'A B^{+}_{out}} \\
  0& \rho_{22}^{'A B^{+}_{out}} & -\rho_{23} ^{'A B^{+}_{out}}& 0\\
  0& -\rho_{23}^{'A B^{+}_{out}} & \rho_{33}^{'A B^{+}_{out}} & 0\\
 -\rho_{14}^{'A B^{+}_{out}} & 0 &0  &\rho_{44}^{'A B^{+}_{out}}
\end{array} \right),\nonumber\\
\end{eqnarray}
where
\begin{eqnarray*}
&&\rho_{11}^{'A B^{+}_{out}}=\frac{(1-ft)}{z_{1}} [\cos^{2}r \rho_{11}+\cos^{2}r q^{2}_{L} \rho_{22}\nonumber\\
&&+p(\cos^{2}r \rho_{33}+\cos^{2}r q^{2}_{L} \rho_{44})],  \nonumber\\
&&\rho_{22}^{'A B^{+}_{out}}=\frac{(1-ft)}{z_{1}} [\sin^{2}r \rho_{11}+\cos^{2}r q^{2}_{R} \rho_{22}+\sin^{2}r \rho_{22} \nonumber\\
&&+p(\sin^{2}r \rho_{33}+\cos^{2}r q^{2}_{R} \rho_{44}+\sin^{2}r \rho_{44})],  \nonumber\\
&&\rho_{33}^{'A B^{+}_{out}}=\frac{ft}{z_{1}} [(1-p) (\cos^{2}r \rho_{33}+\cos^{2}r q^{2}_{L} \rho_{44})], \nonumber\\
&&\rho_{44}^{'A B^{+}_{out}}=\frac{ft}{z_{1}} [(1-p) (\sin^{2}r \rho_{33}+\cos^{2}r q^{2}_{R} \rho_{44}+\sin^{2}r \rho_{44})], \nonumber\\
&&\rho_{14}^{'A B^{+}_{out}}=\frac{\sqrt{1-ft} \sqrt{ft}}{z_{1}} \sqrt{1-p} \cos r q_{R} \rho_{14}, \nonumber\\
&&\rho_{23}^{'A B^{+}_{out}}=\frac{\sqrt{1-ft} \sqrt{ft}}{z_{1}}\sqrt{1-p} \cos r q_{R} \rho_{23},\nonumber\\
\end{eqnarray*}
\begin{eqnarray}
&&z_{1}=ft (1-p) (\rho_{33}+\rho_{44})\nonumber\\
&&-(ft-1)(\rho_{11}+\rho_{22}+p \rho_{33}+p \rho_{44}),
\end{eqnarray}
with $\cos r$ and $\sin r$ defined in Sec. 2. We assume that the state $\rho_{f t}^{'A B^{+}_{out}}$ satisfies the condition $\rho_{22}^{'A B^{+}_{out}}+\rho_{33}^{'A B^{+}_{out}} \ge 1/2$. Thus, by employing the Eq.(13), the fully entangled fraction can be derived as
\begin{eqnarray}
&&f(\rho_{f t}^{'A B^{+}_{out}})=\frac{1}{2 z_{1}} ((1-ft) [\sin^{2}r \rho_{11}+\cos^{2}r q^{2}_{R} \rho_{22}\nonumber\\
&&+\sin^{2}r \rho_{22} +p(\sin^{2}r \rho_{33}+\cos^{2}r q^{2}_{R} \rho_{44}+\sin^{2}r \rho_{44})]\nonumber\\
&&+ft [(1-p) (\cos^{2}r \rho_{33}+\cos^{2}r q^{2}_{L} \rho_{44})]\nonumber\\
&&+2 \sqrt{1-ft} \sqrt{ft} \sqrt{1-p} \cos r q_{R} \rho_{23}).
\end{eqnarray}

The change of the fully entangled fraction $f(\rho_{f t}^{'A B^{+}_{out}})$ related to the dilaton can be given by 
\begin{eqnarray*}
\Delta_{\alpha} f(\rho_{X}^{'A B^{+}_{out}}(\alpha_{0}))=f(\rho_{X}^{A B^{+}_{out}})|_{(\alpha=\alpha_{0})}-f(\rho_{X}^{A B^{+}_{out}})|_{(\alpha=0)}  . 
\end{eqnarray*}

The derivative of $f(\rho_{X}^{'A B^{+}_{out}})$ with respect to the dilaton $\alpha$ can be expressed as $\frac{\partial f(\rho_{X}^{'A B^{+}_{out}})}{\partial \alpha}$. We can find that $\frac{\partial f(\rho_{X}^{'A B^{+}_{out}})}{\partial \alpha} \mid_{\alpha_{0}}>0$ denotes $\Delta_{\alpha} f(\rho_{X}^{'A B^{+}_{out}}(\alpha_{0}))>0$ , and $\frac{\partial f(\rho_{X}^{'A B^{+}_{out}})}{\partial \alpha} \mid_{\alpha_{0}}<0$ denotes $\Delta_{\alpha} f(\rho_{X}^{'A B^{+}_{out}}(\alpha_{0}))<0$.

\begin{figure}
\begin{center}
\includegraphics[width=6cm]{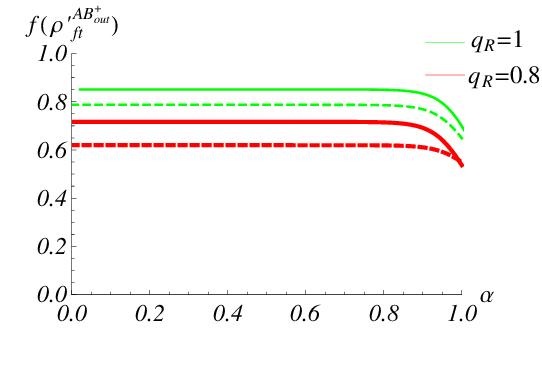}
\caption{\label{fig:fig4} 
The influence of the local filtering operation on the fully entangled fraction $f(\rho_{X}^{' A B^{+}_{out}})$ is plotted as a function of the dilaton parameter $\alpha$. The initial parameters are set as: $\rho_{11}=\rho_{44}=\rho_{14}=0$ and $\rho_{22}=\rho_{33}=\rho_{23}=\frac{1}{2}$. The green dashed line denotes $q_{R}=1$, $p=0.4$ and $ft=0$ (without local filtering operation). The green solid line corresponds to $q_{R}=1$, $p=0.4$ and $ft=0.7$.  The red dashed line describes $q_{R}=0.8$, $p=0.4$ and $ft=0$ (without local filtering operation). The red solid line refers to $q_{R}=0.8$, $p=0.4$ and $ft=0.75$. }
\end{center}
\end{figure}

In Fig. 4, the fully entangled fraction $f(\rho_{f t}^{'A B^{+}_{out}})$ is plotted as a function of the dilaton parameter $\alpha$ for different values of $q_{R}$ under decoherence, where Alice performs a local filtering operation on her particle. By comparing the solid line and the dashed line in Fig.4, it is found that in the case where Alice interacts with the environment, regardless of the value of $q_{R}$ for the Unruh mode, we can use the local filtering operation to increase the value of the fully entangled fraction $f(\rho_{f t}^{'A B^{+}_{out}})$. This means that we can utilize the local filtering operation to suppress the effect of decoherence, thereby enhancing the value of the fully entangled fraction. This is very valuable for the effective execution of quantum teleportation tasks in open systems within the context of a dilaton black hole background.

\begin{figure}
\begin{center}
\includegraphics[width=6cm]{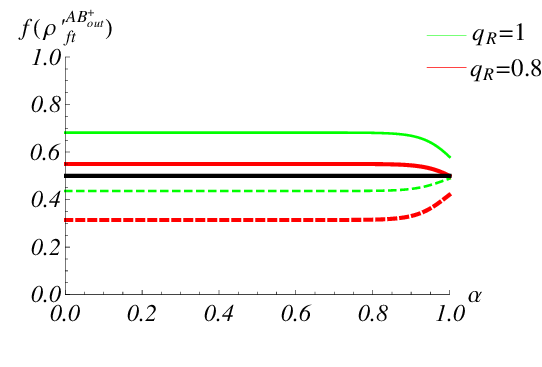}
\caption{\label{fig:fig5} 
The influence of the local filtering operation on the fully entangled fraction $f(\rho_{X}^{' A B^{+}_{out}})$ is plotted as a function of the dilaton parameter $\alpha$. The initial parameters are set as: $\rho_{11}=\sqrt{2}-1$,  $\rho_{22}=\frac{1}{2},$ $\rho_{33}=\frac{3-2 \sqrt{2}}{2}$, $\rho_{14}=\rho_{44}=0$ and $\rho_{23}=\frac{\sqrt{2}-1}{2}$. The green dashed line denotes $q_{R}=1$, $p=0.4$ and $ft=0$ (without local filtering operation). The green solid line corresponds to $q_{R}=1$, $p=0.4$ and $ft=0.9$.
The red dashed line describes $q_{R}=0.8$, $p=0.4$ and $ft=0$ (without local filtering operation). The red solid line refers to $q_{R}=0.8$, $p=0.4$ and $ft=0.9$. The black solid line denotes $f(\rho_{X}^{' A B^{+}_{out}})=0.5$. }
\end{center}
\end{figure}

In Fig. 5, the fully entangled fraction $f(\rho_{f t}^{'A B^{+}_{out}})$ is shown as a function of the dilaton parameter $\alpha$ with a local filtering operation in the context of Alice passing through the AD channel. From the dashed lines of Fig.5, we can see that when Alice interacts with the environment, the fully entangled fraction $f(\rho_{f t}^{'A B^{+}_{out}})$ decreases as $q_{R}$ declines. In particular, by comparing the solid black line and the dashed line in the figure, it can be observed that the value of the fully entangled fraction is less than $0.5$, which means that the fully entangled fraction $f(\rho_{f t}^{'A B^{+}_{out}})$ remains in the classical region as the dilaton parameter $\alpha$ increases. However, since the quantum region $(f(\rho_{f t}^{'A B^{+}_{out}})>0.5)$ can provide better fidelity of quantum teleportation than classical communication, this result is very unfavorable for the effective implementation of quantum teleportation. From the green solid line and red solid line, it is clear that as the dilaton parameter $\alpha$ increases, both the green solid line and the red solid line exceed $0.5$, suggesting that local filtering operation transforms the fully entangled fraction $f(\rho_{f t}^{'A B^{+}_{out}})$ from the classical region to the quantum region, which is very favorable for the effective execution of quantum teleportation. This result has been observed for the first time and is beneficial for the effective implementation of quantum teleportation in the context of a dilaton black hole under decoherence.

\begin{figure}
\begin{center}
\includegraphics[width=6cm]{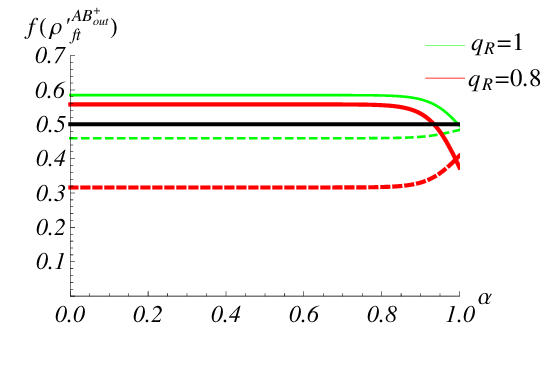}
\caption{\label{fig:fig6} 
The influence of the local filtering operation on the fully entangled fraction $f(\rho_{X}^{' A B^{+}_{out}})$ is plotted as a function of the dilaton parameter $\alpha$. The initial parameters are set as: $\rho_{11}=\frac{\sqrt{2}-1}{2}$, $ \rho_{22}=\frac{\sqrt{2}}{2}$, $\rho_{33}=\frac{3-2 \sqrt{2}}{2}$,\, $\rho_{14}=\rho_{44}=0$ and $\rho_{23}=\frac{\sqrt{2}-1}{4}$. The green dashed line denotes $q_{R}=1$, $p=0.4$ and $ft=0$ (without local filtering operation). The green solid line corresponds to $q_{R}=1$, $p=0.4$ and $ft=0.9$. The red dashed line describes $q_{R}=0.8$, $p=0.4$ and $ft=0$ (without local filtering operation). The red solid line refers to $q_{R}=0.8$, $p=0.4$ and $ft=0.98$. The black solid line denotes $f(\rho_{X}^{' A B^{+}_{out}})=0.5$. }
\end{center}
\end{figure}

In Fig.6, we plot the fully entangled fraction $f(\rho_{f t}^{'A B^{+}_{out}})$ as a function of the dilaton parameter $\alpha$ for different values of initial state under decoherence, while a local filtering operation is applied to Alice. From the Figs.5 and 6, it can be observed that even under the different initial states, the fully entangled fraction $f(\rho_{f t}^{'A B^{+}_{out}})$ is always less than $0.5$, which indicates that the fully entangled fraction $f(\rho_{f t}^{'A B^{+}_{out}})$ stays in the classical region as the dilaton parameter $\alpha$ increases. It is worth noting that after we perform the local filtering operation on Alice, the value of the fully entangled fraction $f(\rho_{f t}^{'A B^{+}_{out}})$ increases above $0.5$. This means that the local filtering operation can convert the fully entangled fraction from the classical region to the quantum region even under the different initial states, that is, this provides better fidelity than classic communication, which is crucial for the effective execution of quantum information tasks.

From Figs. 4, 5, and 6, it is obvious that when the system interacts with the environment, the fidelity of quantum teleportation can be enhanced by applying a local filtering operation on Alice, irrespective of the values of the Unruh modes and the initial states. In particular, when the fully entangled fraction lies within the classical region, the local filtering operation can be employed to shift the fully entangled fraction from the classical region to the quantum region, thus achieving greater fidelity than classical communication.

\section{CONCLUSION}

In this paper, we investigate the influence of the dilaton effect on a general quantum teleportation protocol between users beyond the single-mode approximation in the background of a GHS dilaton black hole under decoherence. Alice and Bob initially share an X-type state and employ a standard teleportation scheme (STS) to transfer the unknown state from Alice to Bob. We consider the scenario where the sender, Alice, remains stationary in an asymptotically flat region and is coupled with the environment, while the receiver, Bob, moves with uniform acceleration toward the event horizon of the dilaton black hole and hovers nearby. It is found that as the dilaton parameter $\alpha$ increases, the fidelity of quantum teleportation can either decrease or increase, depending on the choice of the initial state and the decoherence parameter $p$. This indicates that the dilaton effect of the black hole and environmental decoherence positively influences the fidelity of teleportation; specifically, the dilaton effect can enhance and create net fidelity in quantum teleportation when Alice’s particle interacts with an amplitude damping channel. These results differ from previous research findings \cite{aa26aa,a32a} and may challenge the long-standing belief that the effects of black holes and decoherence can only destroy the fidelity of quantum teleportation, which may provide valuable insights for a more comprehensive understanding of the effects of the dilaton black hole and decoherence on the effective execution of quantum information tasks in real quantum systems. 

Furthermore, it is worth noting that Wu et al. \cite{a55a} recently explored the impact of the Unruh effect on 1-3 quantum entanglement within the framework of entangled tetrapartite Unruh-DeWitt detectors. They observed that the quantum entanglement of the tetrapartite W state first decreases to a minimum and then increases to a fixed value with increasing acceleration. This demonstrates that the Unruh effect can both degrade and enhance entanglement. In other words, under specific conditions, the Unruh effect is capable of enhancing quantum entanglement; specifically, quantum entanglement can be extracted from the quantum field to the detectors. These results indicate that relativistic effects can increase quantum entanglement, which is similar to our finding that gravitational effects can enhance the fidelity of quantum teleportation.  Both our results and those of Wu et al. challenge the conventional paradigm that relativistic and gravitational effects only degrade quantum resources. They reveal the dual capacity of relativistic and gravitational effects, which can either diminish or enhance quantum properties under varying conditions.  These findings provide new insights for understanding gravitational and relativistic effects in relativistic quantum information research.

Finally, we find that the fidelity of quantum teleportation depends on the choice of Unruh modes. It is shown that the Unruh mode with $q_{R}=1$ is always optimal, achieving the best transmission of an unknown pure state to Bob. Finally, we explore the feasibility of increasing the fidelity of quantum teleportation in dilaton spacetime under decoherence by employing a local filtering operation. It is demonstrated that when the system is in an open environment, we can improve the fidelity of quantum teleportation by performing the local filtering operation on Alice, regardless of the values of the Unruh modes and the initial states. Specifically, some previous studies \cite{a10a,a11a} have shown that in curved spacetime, the fully entangled fraction stays in the classical region under some cases, which means that the STS scheme cannot provide better fidelity than classical communication. However, our results demonstrate that when the value of the fully entangled fraction is in the classical region, we can use the local filtering operation to transform the fully entangled fraction from the classical region to the quantum region, thereby achieving better fidelity than classical communication. These results may provide a new perspective for finding an experimental scheme to effectively implement quantum teleportation in the context of dilaton black holes under decoherence.

\begin{acknowledgments}
This project was supported by the National Natural Science Foundation of China (Grant Nos. 12265007 and 11364006).
\end{acknowledgments}

\appendix
\section{The detailed calculation process of Eq. (15)}
In this appendix, we briefly outline the steps to derive Eq. (15). An X-type state is initially shared by Alice and Bob at an asymptotically flat region, which is described in Eq. (10). Then, we let Bob hover near the event horizon of the black hole. According to Eqs. (8) and (9), we can rewrite Eq. (10) as 
\begin{eqnarray*}
 &&\rho_{X}=\left | \Psi  \right \rangle _{AB^{+}B^{-}b^{+}b^{-}} \left\langle\Psi\right| \nonumber\\
 && =\rho_{11}  [  | 0  \rangle ( \cos^{2} r \left | 0000 \right \rangle-\sin r \cos r \left | 0011 \right \rangle +\sin r \cos r \left | 1100 \right \rangle\nonumber\\
 &&-\sin^{2} r\left | 1111 \right \rangle )   \otimes \left\langle 0 \right| (\cos^{2} r \left\langle 0000 \right|-\sin r \cos r\left\langle 0011 \right| \nonumber\\
  &&+ \sin r \cos r  \left\langle 1100 \right| -\sin^{2} r\left\langle 1111 \right | ) ]-\rho_{14}  [  | 0  \rangle ( \cos^{2} r \left | 0000 \right \rangle
\nonumber\\
  &&-\sin r \cos r \left | 0011 \right \rangle +\sin r \cos r \left | 1100 \right \rangle-\sin^{2} r\left | 1111 \right \rangle) \nonumber\\
  &&  \otimes \left\langle 1 \right| (q_{R} \cos r \left\langle 1000 \right|-q_{R} \sin r \left\langle 1011 \right| +q_{L} \sin r   \left\langle 1101 \right| \nonumber\\
 &&+q_{L} \cos r\left\langle 0001  \right|    ) ]+\rho_{22}  [  | 0  \rangle ( q_{R} \cos r\left | 1000 \right \rangle-q_{R} \sin r\left | 1011 \right \rangle \nonumber\\
 &&+q_{L} \sin r \left | 1101 \right \rangle\ + 
 q_{L} \cos r\left | 0001 \right \rangle) \otimes \left\langle 0 \right| (q_{R} \cos r \left\langle 1000 \right|\nonumber\\
  &&  -q_{R} \sin r \left\langle 1011 \right| +q_{L} \sin r   \left\langle 1101 \right| +q_{L} \cos r\left\langle 0001 \right | ) ]
\nonumber\\
&&-\rho_{23}  [  | 0  \rangle ( q_{R} \cos r\left | 1000 \right \rangle-q_{R} \sin r\left | 1011 \right \rangle+q_{L} \sin r \left | 1101 \right \rangle\ \nonumber\\
  && + 
 q_{L} \cos r \left | 0001 \right \rangle) \otimes \left\langle 1 \right| (\cos^{2} r \left\langle 0000 \right|-\sin r \cos r\left\langle 0011 \right| \nonumber\\
  &&+ \sin r \cos r  \left\langle 1100 \right| -\sin^{2} r\left\langle 1111 \right| ) ]-\rho_{23}  [  | 1  \rangle ( \cos^{2} r \left | 0000 \right \rangle
\nonumber\\
&&-\sin r \cos r \left | 0011 \right \rangle +\sin r \cos r \left | 1100 \right \rangle-\sin^{2} r\left | 1111 \right \rangle ) \nonumber\\
   &&  \otimes\left\langle 0 \right| (q_{R} \cos r \left\langle 1000 \right|-q_{R} \sin r \left\langle 1011 \right|+q_{L} \sin r   \left\langle 1101 \right|  \nonumber\\
   &&+q_{L} \cos r\left\langle 0001 \right | ) ]+\rho_{33}  [  | 1  \rangle ( \cos^{2} r \left | 0000 \right \rangle-\sin r \cos r \left | 0011 \right \rangle \nonumber\\
          &&+\sin r \cos r \left | 1100 \right \rangle-\sin^{2} r\left | 1111 \right \rangle )  \otimes \left\langle 1 \right| (\cos^{2} r \left\langle 0000 \right|\nonumber\\
  && -\sin r \cos r\left\langle 0011 \right| + \sin r \cos r  \left\langle 1100 \right| -\sin^{2} r\left\langle 1111 \right| ) ]
\nonumber\\
  &&-\rho_{14}  [  | 1  \rangle ( q_{R} \cos r\left | 1000 \right \rangle-q_{R} \sin r\left | 1011 \right \rangle +q_{L} \sin r \left | 1101 \right \rangle\ \nonumber\\
  && +  q_{L} \cos r\left | 0001 \right \rangle)  \otimes\left\langle 0 \right| (\cos^{2} r \left\langle 0000 \right|-\sin r \cos r\left\langle 0011 \right| \nonumber\\
 &&+ \sin r \cos r  \left\langle 1100 \right| -\sin^{2} r\left\langle 1111 \right | ) ]+\rho_{44}  [  | 1  \rangle ( q_{R} \cos r\left | 1000 \right \rangle
\nonumber\\
 &&-q_{R} \sin r\left | 1011 \right \rangle+ q_{L} \sin r \left | 1101 \right \rangle\ + 
 q_{L} \cos r\left | 0001 \right \rangle) \nonumber\\
 \end{eqnarray*}
 \begin{eqnarray}
 &&  \otimes \left\langle 1 \right| (q_{R} \cos r \left\langle 1000 \right|-q_{R} \sin r \left\langle 1011 \right| \nonumber\\
 &&+q_{L} \sin r   \left\langle 1101 \right| +q_{L} \cos r\left\langle 0001 \right | ) ].
\end{eqnarray}

Here, we analyze the scenario in which Bob detects a thermal Fermi-Dirac particle distribution using his excited detector. Since Bob cannot access the modes inside the black hole's event horizon, we trace over these inaccessible modes and derive the reduced density matrix $\rho_{X}^{A B_{\text{out}}}$ as

\begin{eqnarray}
&&\rho_{X}^{A B_{out}}=Tr_{B^{-}b^{+}b^{-}}\left ( \left | \Psi  \right \rangle _{AB^{+}B^{-}b^{+}b^{-}} \left\langle\Psi\right|         \right ) \nonumber\\
&&=\left( \begin{array}{cccc}
 \rho_{11}^{A B_{out}} & 0 & 0 &-\rho_{14}^{A B_{out}} \\
  0& \rho_{22}^{A B_{out}} & -\rho_{23} ^{A B_{out}}& 0\\
  0& -\rho_{23}^{A B_{out}} & \rho_{33}^{A B_{out}} & 0\\
 -\rho_{14}^{A B_{out}} & 0 &0  &\rho_{44}^{A B_{out}}
\end{array} \right),
\end{eqnarray}
where
\begin{eqnarray}
\rho_{11}^{A B_{out}}&=&\cos^{2}r (\rho_{11}+q^{2}_{L} \rho_{22}),  \nonumber\\
\rho_{22}^{A B_{out}}&=&\sin^{2}r \rho_{11}+\cos^{2}r q^{2}_{R} \rho_{22}+\sin^{2}r \rho_{22} ,  \nonumber\\
\rho_{33}^{A B_{out}}&=&\cos^{2}r \rho_{33}+\cos^{2}r q^{2}_{L} \rho_{44}, \nonumber\\
\rho_{44}^{A B_{out}}&=&\sin^{2}r \rho_{33}+\cos^{2}r q^{2}_{R} \rho_{44}+\sin^{2}r \rho_{44}, \nonumber\\
\rho_{14}^{A B_{out}}&=& \cos r q_{R} \rho_{14}, \rho_{23}^{A B}= \cos r q_{R} \rho_{23}.
\end{eqnarray}

In the following, we consider the scenario in which Alice undergoes an amplitude damping (AD) channel, as described by Eq. (14). By virtue of the Eq. (14), we can rewrite the Eq. (A2) as

\begin{widetext}
\begin{eqnarray}
&&\rho_{X}^{AA_{E}B_{out}}= \rho^{AA_{E}B_{out}}_{11}|000\rangle_{{AA_{E}B_{out}} }\langle 000|-\sqrt{p}\rho^{AA_{E}B_{out}}_{14}|000\rangle_{{AA_{E}B_{out}} }\langle 011|-\sqrt{1-p}\rho^{AA_{E}B_{out}}_{14}|000\rangle_{{AA_{E}B_{out}} }\langle 101|\nonumber\\
&&+\rho^{A A_{E}B_{out}}_{22}|001\rangle_{{AA_{E}B_{out}} }\langle 001|-\sqrt{p}\rho^{AA_{E}B_{out}}_{23}|001\rangle_{{AA_{E}B_{out}} }\langle 010|-\sqrt{1-p}\rho^{AA_{E}B_{out}}_{23}|001\rangle_{{AA_{E}B_{out}} }\langle 100|\nonumber\\
&&-\sqrt{p}\rho^{A A_{E}B_{out}}_{23}|010\rangle_{{AA_{E}B_{out}} }\langle 001|+p\rho^{AA_{E}B_{out}}_{33}|010 \rangle_{{AA_{E}B_{out}} }\langle 010|+\sqrt{p(1-p)}\rho^{A A_{E}B_{out}}_{33}|010\rangle_{{AA_{E}B_{out}} }\langle 100|\nonumber\\
&&-\sqrt{p}\rho^{AA_{E}B_{out}}_{14}|011 \rangle_{{AA_{E}B_{out}} }\langle 000|+p\rho^{AA_{E}B_{out}}_{44}|011 \rangle_{{AA_{E}B_{out}} }\langle 011|+\sqrt{p(1-p)}\rho^{AA_{E}B_{out}}_{44}|011 \rangle_{{AA_{E}B_{out}} }\langle 101|\nonumber\\
&&-\sqrt{1-p}\rho^{AA_{E}B_{out}}_{23}|100 \rangle_{{AA_{E}B_{out}} }\langle 001|+\sqrt{p(1-p)}\rho^{AA_{E} B_{out}}_{33}|100 \rangle_{{AA_{E}B_{out}} }\langle 010|+(1-p)\rho^{AA_{E}B_{out}}_{33}|100 \rangle_{{AA_{E}B_{out}} }\langle 100|\nonumber\\
&&-\sqrt{1-p}\rho^{AA_{E}B_{out}}_{14}|101 \rangle_{{AA_{E}B_{out}} }\langle 000|+\sqrt{p(1-p)}\rho^{A A_{E}B_{out}}_{44}|101 \rangle_{{AA_{E}B_{out}} }\langle 011|+\sqrt{1-p}\rho^{AA_{E}B_{out}}_{44}|101 \rangle_{{AA_{E}B_{out}} }\langle 101|,\nonumber\\
\end{eqnarray}
with
\begin{eqnarray}
\rho^{AA_{E}B_{out}}_{11}&=&\rho^{AB_{out}}_{11}=\cos^{2}r (\rho_{11}+q^{2}_{L} \rho_{22}),  
\rho^{AA_{E}B_{out}}_{14}=\rho^{AB_{out}}_{14}=\cos r q_{R} \rho_{14}, 
\nonumber\\
\rho^{AA_{E}B_{out}}_{22}&=&\rho^{AB_{out}}_{22}=\sin^{2}r \rho_{11}+\cos^{2}r q^{2}_{R} \rho_{22}+\sin^{2}r \rho_{22}, 
\nonumber\\
\rho^{AA_{E}B_{out}}_{33}&=&\rho^{AB_{out}}_{33}=\cos^{2}r \rho_{33}+\cos^{2}r q^{2}_{L} \rho_{44},\rho^{AA_{E}B_{out}}_{23}=\rho^{AB_{out}}_{23}=\cos r q_{R} \rho_{23},   \nonumber\\
\rho^{AA_{E}B_{out}}_{44}&=&\rho^{AB_{out}}_{44}=\sin^{2}r \rho_{33}+\cos^{2}r q^{2}_{R} \rho_{44}+\sin^{2}r \rho_{44}. 
\end{eqnarray}    
\end{widetext}

By tracing over the degrees of freedom of the environment, the density matrix $\rho_{X}^{A B^{+}_{\text{out}}}$ can be obtained, as defined in Eq. (15).

\end{document}